\pdfoutput=1

\documentclass[sigconf,screen=true,natbib,9pt]{acmart}

\copyrightyear{2018} 
\acmYear{2018} 
\setcopyright{acmlicensed}
\acmConference[CIKM '18]{The 27th ACM International Conference on Information and Knowledge Management}{October 22--26, 2018}{Torino, Italy}
\acmBooktitle{The 27th ACM International Conference on Information and Knowledge Management (CIKM '18), October 22--26, 2018, Torino, Italy}
\acmPrice{15.00}
\acmDOI{10.1145/3269206.3271686}
\acmISBN{978-1-4503-6014-2/18/10}

\usepackage{booktabs} 
\usepackage{graphicx}
\usepackage{url}
\usepackage{microtype}
\usepackage{tabularx,amsmath}
\usepackage{amssymb}
\usepackage{multirow}
\usepackage{color}
\usepackage[normalem]{ulem}
\usepackage[english]{babel}
\usepackage{colortbl}
\usepackage[noend]{algorithmic}
\usepackage{algorithm}
\usepackage{paralist}
\usepackage{enumerate}
\usepackage{enumitem}
\usepackage{acronym}
\usepackage[skip=0pt]{caption}
\usepackage{setspace}
\usepackage{balance}
\usepackage{subfig}
\usepackage[np, autolanguage]{numprint}

\setcitestyle{square,comma,numbers,sort&compress,sectionbib}

\newbox\itembox
\def\itemlistlabel#1{#1\hfill}
\def\itemlist#1{\setbox\itembox=\hbox{#1}%
                \list{}{\labelwidth\wd\itembox
                             \leftmargin\labelwidth
                             \advance\leftmargin by\itemindent
                             \advance\leftmargin by\labelsep
                             \let\makelabel\itemlistlabel}}

\newcommand{\OtherDocuments}{\{\ldots\}}

\newcommand{\enkelop}{$^{\vartriangle}$}
\newcommand{\dubbelop}{$^{\blacktriangle}$}
\newcommand{\enkelneer}{$^{\triangledown}$}
\newcommand{\dubbelneer}{$^{\blacktriangledown}$}

\acrodef{IR}{Information Retrieval}
\acrodef{LTR}{Learning to Rank}
\acrodef{OLTR}{Online Learning to Rank}
\acrodef{DBGD}{Dueling Bandit Gradient Descent}
\acrodef{MGD}{Multileave Gradient Descent}
\acrodef{C-MGD}{Cascading Multileave Gradient Descent}
\acrodef{RL}{Reinforcement Learning}
\acrodef{PL}{Plackett-Luce}

\acrodef{PUB-Rank}{Pairwise Unbiased Ranker Optimization}
\acrodef{PUGD}{Pairwise Unbiased Gradient Descent}
\acrodef{PDGD}{Pairwise Differentiable Gradient Descent}
\newcommand{\OurMethod}{PDGD}

\title{Beyond Linear Models: A Novel Differentiable Unbiased Learning to Rank Method for the Online Setting}
\title{Differentiable Unbiased Online Learning to Rank}

\author{Harrie Oosterhuis}
\orcid{}
\affiliation{%
\institution{University of Amsterdam}
}
\email{oosterhuis@uva.nl}

\author{Maarten de Rijke}
\orcid{0000-0002-1086-0202}
\affiliation{%
\institution{University of Amsterdam}
}
\email{derijke@uva.nl}

\begin{document}

\begin{abstract}

\acf{OLTR} methods optimize rankers based on user interactions.
State-of-the-art \acs{OLTR} methods are built specifically for linear models. 
Their approaches do not extend well to non-linear models such as neural networks.
We introduce an entirely novel approach to \acs{OLTR} that constructs a weighted differentiable pairwise loss after each interaction: \acf{\OurMethod}.
\ac{\OurMethod} breaks away from the traditional approach that relies on interleaving or multileaving and extensive sampling of models to estimate gradients.
Instead, its gradient is based on inferring preferences between document pairs from user clicks and can optimize any differentiable model.
We prove that the gradient of \acs{\OurMethod} is unbiased w.r.t.\ user document pair preferences.
Our experiments on the largest publicly available \ac{LTR} datasets show considerable and significant improvements under all levels of interaction noise.
\acs{\OurMethod} outperforms existing \ac{OLTR} methods both in terms of learning speed as well as final convergence.
Furthermore, unlike previous \ac{OLTR} methods, \ac{\OurMethod} also allows for non-linear models to be optimized effectively.
Our results show that using a neural network leads to even better performance at convergence than a linear model.
In summary, \acs{\OurMethod} is an efficient and unbiased \ac{OLTR} approach that provides a better user experience than previously possible.
\end{abstract}

%
%
%

\maketitle

\acresetall


\section{Introduction}
\label{sec:intro}

In order to benefit from unprecedented volumes of content, users rely on ranking systems to provide them with the content of their liking.
\ac{LTR} in \ac{IR} concerns methods that optimize ranking models so that they order documents according to user preferences.
In web search engines such models combine hundreds of signals to rank web-pages according to their relevance to user queries \cite{liu2009learning}.
Similarly, ranking models are a vital part of recommender systems where there is no explicit search intent \cite{karatzoglou2013learning}.
\ac{LTR} is also prevalent in settings where other content is ranked, e.g., videos~\citep{chelaru2014useful}, products~\citep{karmaker2017application}, conversations~\citep{DBLP:conf/chiir/RadlinskiC17} or personal documents~\citep{wang2016learning}.

Traditionally, \ac{LTR} has been applied in the \emph{offline} setting where a dataset with annotated query-document pairs is available.
Here, the model is optimized to rank documents according to the relevance annotations, which are based on the judgements of human annotators.
Over time the limitations of this supervised approach have become apparent: annotated sets are expensive and time-consuming to create~\citep{letor, Chapelle2011}; when personal documents are involved such a dataset would breach privacy~\citep{wang2016learning}; the relevance of documents to queries can change over time, like in a news search engine~\citep{dumais-web-2010,lefortier-online-2014}; and judgements of raters are not necessarily aligned with the actual users \citep{sanderson2010}.

In order to overcome the issues with annotated datasets, previous work in \ac{LTR} has looked into learning from user interactions.
Work along these lines can be divided into \emph{approaches that learn from historical interactions}, i.e., in the form of interaction logs~\citep{Joachims2002}, and \emph{approaches that learn in an online setting}~\citep{yue09:inter}.
The latter regard methods that determine what to display to the user at each impression, and then immediately learn from observed user interactions and update their behavior accordingly.
This online approach has the advantage that it does not require an existing ranker of decent quality, and thus can handle cold-start situations.
Additionally, it is more responsive to the user by updating continuously and instantly, therefore allowing for a better experience.
However, it is important that an online method can handle biases that come with user behavior:
for instance, the observed interactions only take place with the displayed results, i.e., there is selection bias, and are more likely to occur with higher ranked items, i.e., there is position bias.
Accordingly, a method should learn user preferences w.r.t.\ document relevance, and be robust to the forms of noise and bias present in the online setting.
Overall, the online \ac{LTR} approach promises to learn ranking models that are in line with user preferences, in a responsive matter, reaching good performance from few interactions, even in cold-start situations.

Despite these highly beneficial properties, previous work in \ac{OLTR} has only considered linear models \cite{schuth2016mgd, hofmann12:balancing, yue09:inter} or trivial variants thereof~\cite{oosterhuis2017balancing}.
The reason for this is that existing work in \ac{OLTR} has worked with the \ac{DBGD} algorithm \cite{yue09:inter} as a basis.
While very influential and effective, we identify two main problems with the gradient estimation of the \ac{DBGD} algorithm:
\begin{enumerate}[align=left,leftmargin=*,topsep=1pt]
\item Gradient estimation is based on sampling model variants from a unit circle around the current model. 
This concept does not extend well to non-linear models. Computing rankings for variants is also computationally costly for larger complex models.
\item It uses online evaluation methods, i.e., interleaving or multileaving, to determine the gradient direction from the resulting set of models.
However, these evaluation methods are designed for finding preferences between ranking systems, not (primarily) for determining how a model should be updated.
\end{enumerate}
As an alternative we introduce \acfi{\OurMethod}\acused{\OurMethod}, the first unbiased \ac{OLTR} method that is applicable to any differentiable ranking model.
\ac{\OurMethod} infers pairwise document preferences from user interactions and constructs an unbiased gradient after each user impression.
In addition, \ac{\OurMethod} does not rely on sampling models for exploration, but instead models rankings as probability distributions over documents.
Therefore, it allows the \ac{OLTR} model to be very certain for specific queries and perform less exploration in those cases, while being much more explorative in other, uncertain cases.
Our results show that, consequently, \ac{\OurMethod} provides significant and considerable improvements over previous \ac{OLTR} methods.
This indicates that its gradient estimation is more in line with the preferences to be learned.

In this work, we answer the following three research questions:
 \begin{enumerate}[align=left, label={\bf RQ\arabic*},leftmargin=*,topsep=1pt]
    \item Does using \ac{\OurMethod} result in significantly better performance than the current state-of-the-art \acl{MGD}?\label{rq:performance}
    \item Is the gradient estimation of \ac{\OurMethod} unbiased? \label{rq:unbiased}
    \item Is \ac{\OurMethod} capable of effectively optimizing different types of ranking models? \label{rq:nonlinear}
\end{enumerate}
To facilitate replicability and repeatability of our findings, we provide open source implementations of \ac{\OurMethod} and our experiments under the permissive MIT open-source license.\footnote{https://github.com/HarrieO/OnlineLearningToRank} 


\section{Related Work}
\label{sec:relatedwork}

%
\subsection{Learning to rank}
\ac{LTR} can be applied to the offline and online setting.
In the offline setting \ac{LTR} is approached as a supervised problem where the relevance of each query-document pair is known.
Most of the challenges with offline \ac{LTR} come from obtaining annotations.
For instance, gathering annotations is time-consuming and expensive \cite{letor, qin2013introducing, Chapelle2011}.
Furthermore, in privacy sensitive-contexts it would be unethical to annotate items, e.g., for personal emails or documents \cite{wang2016learning}.
Moreover, for personalization problems annotators are unable to judge what specific users would prefer.
Also, (perceived) relevance chances over time, due to cognitive changes on the user's end~\citep{vakkari-changes-2000} or due to changes in document collections~\citep{dumais-web-2010} or the real world~\citep{lefortier-online-2014}.
Finally, annotations are not necessarily aligned with user satisfaction, as judges may interpret queries differently from actual users~\cite{sanderson2010}.
Consequently, the limitations of offline \ac{LTR} have led to an increased interest in alternative approaches to \ac{LTR}.

\subsection{Online learning to rank}
\ac{OLTR} is an attractive alternative as it learns directly from interacting with users~\cite{yue09:inter}.
By doing so it attempts to solve the issues with offline annotations that occur in \ac{LTR}, as user preferences are expected to be better represented by interactions than with offline annotations~\cite{radlinski08:how}.
Unlike methods in the offline setting, \ac{OLTR} algorithms have to simultaneously perform ranking while also optimizing their ranking model.
In other words, an \ac{OLTR} algorithm decides what rankings to display to users, while at the same time learning from the interactions with the presented rankings.
While the potential of learning in the online setting is great, it has its own challenges.
In particular, the main difficulties of the \ac{OLTR} task are \emph{bias} and \emph{noise}.
Any user interaction that does not reflect their true preference is considered noise, this happens frequently e.g., clicks often occur for unexpected reasons~\citep{sanderson2010}.
Bias comes in many forms, for instance, selection bias occurs because interactions only involve displayed documents~\citep{wang2016learning}.
Another common bias is position bias, a consequence from the fact documents at the top of a ranking are more likely to be considered~\citep{yue2010beyond}.
An \ac{OLTR} method should thus take into account the biases that affect user behavior while also being robust to noise, in order to learn the \emph{true} user preferences.

\ac{OLTR} methods can be divided into two groups~\citep{zoghi-online-2017}: \emph{model-based} methods that learn the best ranked list under some model of user interaction with the list~\cite{Radlinski2008,Slivkins2013}, such as a click model~\citep{chuklin-click-2015}, and
\emph{model-free} algorithms that learn the best ranker in a family of rankers~\citep{yue09:inter,hofmann_2013_reusing}.
Model-based methods may have greater statistical efficiency but they give up generality, essentially requiring us to learn a separate model for every query.
For the remainder of this paper, we focus on model-free \ac{OLTR} methods.

\subsection{DBGD and beyond}
State-of-the-art (model-free) \ac{OLTR} approaches learn user preferences by approaching optimization as a dueling bandit problem~\cite{yue09:inter}.
They estimate the gradient of the model w.r.t.\ user satisfaction by comparing the current model to sampled variations of the model.
The original \ac{DBGD} algorithm~\cite{yue09:inter} uses interleaving methods to make these comparisons: at each interaction the rankings of two rankers are combined to create a single result list.
From a large number of clicks on such a combined result list a user preference between the two rankers can reliably be inferred~\cite{hofmann2011probabilistic}.
Conversely, \ac{DBGD} compares its current ranking model to a different slight variation at each impression.
Then, if a click is indicative of a preference for the variation, the current model is slightly updated towards it.
Accordingly, the model of \ac{DBGD} will continuously update itself and oscillate towards an inferred optimum.

Other work in \ac{OLTR} has used \ac{DBGD} as a basis and extended upon it.
Notably, \citet{hofmann_2013_reusing} have proposed a method that guides exploration by only sampling variations that seem promising from historical interaction data.
Unfortunately, while this approach provides faster initial learning, the historical data introduces bias which leads to the quality of the ranking model to steadily decrease over time~\cite{oosterhuis2016probabilistic}.
Alternatively, \citet{schuth2016mgd} introduced \acf{MGD}, this extension replaced the interleaving of \ac{DBGD} with multileaving methods.
In turn the multileaving paradigm is an extension of interleaving where a set of rankers are compared efficiently~\citep{oosterhuis2017sensitive, schuth2015probabilistic, Schuth2014a}.
Conversely, multileaving methods can combine the rankings of more than two rankers and thus infer preferences over a set of rankers from a single click.
\ac{MGD} uses this property to estimate the gradient more effectively by comparing a large number of model variations per user impression~\citep{schuth2016mgd, oosterhuis2016probabilistic}.
As a result, \ac{MGD} requires fewer user interactions to converge on the same level of performance as \ac{DBGD}.
Another alternative approach was considered by \citet{hofmann11:balancing}, who inject the ranking from the current model with randomly sampled documents.
Then, after each user impression, a pairwise loss is constructed from inferred preferences between documents.
This pairwise approach was not found to be more effective than \ac{DBGD}.

Quite remarkably, all existing work in \ac{OLTR} has only considered linear models.
Recently, \citet{oosterhuis2017balancing} recognized that a tradeoff unique to \ac{OLTR} arises when choosing models.
High capacity models such as neural networks~\cite{burges2010ranknet} require more data than simpler models.
On the one hand, this means that high capacity models need more user interactions to reach the same level of performance, thus giving a worse initial user experience.
On the other hand, high capacity models are capable of finding better optima, thus lead to better final convergence and a better long-term user experience.
This dilemma is named the \emph{speed}-\emph{quality} tradeoff, and as a solution a cascade of models can be optimized: combining the initial learning speed of a simple model with the convergence of a complex one.
But there are more reasons why non-linear models have so far been absent from \ac{OLTR}.
Importantly, the \ac{DBGD} was designed for linear models from the ground up; relying on a unit circle to sample model variants and averaging models to estimate the gradient.
Furthermore, the computational cost of maintaining an extensive set of model variants for large and complex models makes this approach very impractical.

Our contribution over the work listed above is an \ac{OLTR} method that is not an extension of \ac{DBGD}, instead it computes differentiable pairwise loss to update its model.
Unlike the existing pairwise approach, our loss function is unbiased and our exploration is performed using the model's confidence over documents.
Finally, we also show that this is the first \ac{OLTR} method to effectively optimize neural networks in the online setting.

\section{Method}
\label{sec:simmgd}

In this section we introduce a novel \ac{OLTR} algorithm: \ac{\OurMethod}.
First, Section~\ref{sec:method:description} describes \ac{\OurMethod} in detail, before Section~\ref{sec:method:unbiased} formalizes and proves the unbiasedness of the method.
Table~\ref{table:notation} lists the notation we use.

\begin{table}[t]
\caption{Main notation used in the paper.}
\label{table:notation} 
\begin{tabular}{m{4.5em}<{\raggedright} m{20em}<{\raggedright}}
    \toprule
 \bf Notation  & \bf Description \\
 \midrule
$d$, $d_k$, $d_l$ & document\\
$\mathbf{d}$ & feature representation of a query-document pair \\
$D$ & set of documents\\
$R$ & ranked list \\
$R^*$ & the reversed pair ranking $R^*(d_k, d_l, R)$ \\
$R_i$ & document placed at rank $i$ \\
$\rho$ & preference pair weighting function \\
$\theta$ & parameters of the ranking model\\
$f_\theta(\cdot)$ & ranking model with parameters $\theta$ \\
$f(\mathbf{d}_k)$ & ranking score for a document from model \\
$\mathit{click}(d)$ & a click on document $d$ \\
$d_k =_\mathit{rel} d_l$ & two documents equally preferred by users \\
$d_k >_\mathit{rel} d_l$ & a user preference between two documents \\
$d_k >_\mathbf{c} d_l$ & document preference inferred from clicks \\
\bottomrule
\end{tabular}
\end{table}

\subsection{\acl{\OurMethod}}
\label{sec:method:description}

\ac{\OurMethod} revolves around optimizing a ranking model $f_{\theta}(\mathbf{d})$ that takes a feature representation of a query-document pair $\mathbf{d}$ as input and outputs a score. 
The aim of the algorithm is to find the parameters $\theta$ so that sorting the documents by their scores in descending order provides the most optimal rankings.
Because this is an online algorithm, the method must first decide what ranking to display to the user, then after the user has interacted with the displayed ranking, it may update $\theta$ accordingly.

\begin{figure}[tb]
\centering

\subfloat[]{
\includegraphics[height=9em]{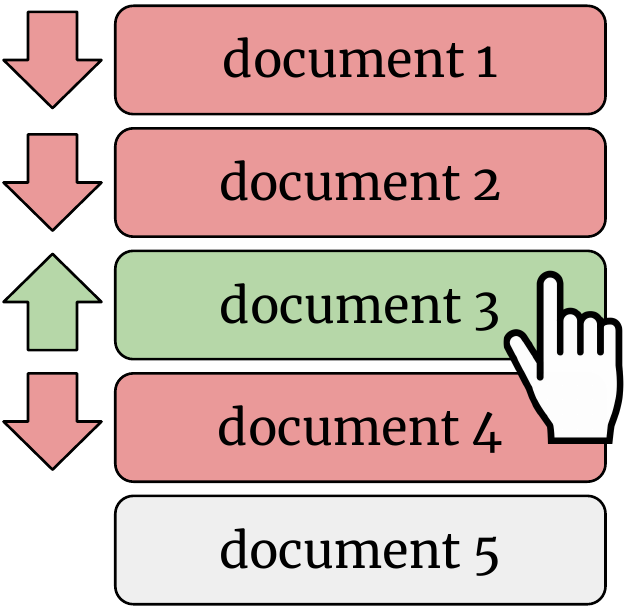}
\label{fig:clickselection}
\hspace{1em} 
}
\subfloat[]{
\includegraphics[height=9em]{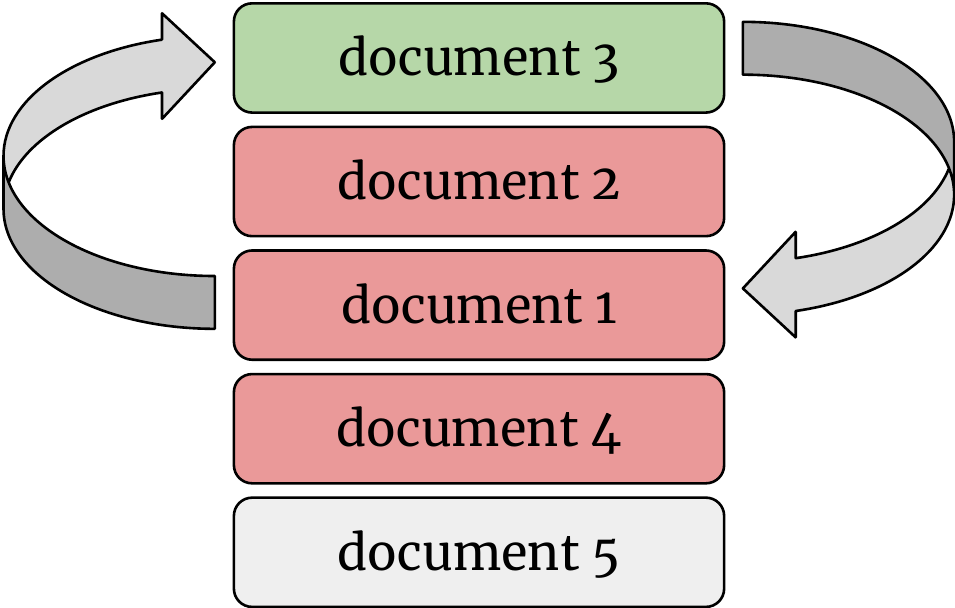}
\label{fig:reversepair}
}
\caption{
Left: a click on a document ranking $R$ and the inferred preferences of $d_3$ over $\{d_1, d_2, d_4\}$.
Right: the reversed pair ranking $R^*(d_1, d_3, R)$ for the document pair $d_1$ and $d_3$.
}
\label{fig:selectionandreverse}
\end{figure}

Unlike previous \ac{OLTR} approaches, \ac{\OurMethod} does not rely on any online evaluation methods. 
Instead, a \ac{PL} model is applied to the ranking function $f_{\theta}(\cdot)$ resulting in a distribution over the document set $D$:
\begin{align}
P(d|D) &= \frac{e^{f_{\theta}(\mathbf{d})}}{\sum_{d' \in D} e^{f_{\theta}(\mathbf{d'})}}.
\label{eq:docprob}
\end{align}
A ranking $R$ to display to the user is then created by sampling from the distribution $k$ times, where after each placement the distribution is renormalized to prevent duplicate placements.
\ac{PL} models have been used before in \ac{LTR}. 
For instance, the ListNet method~\cite{Cao2007} optimizes such a model in the offline setting.
With $R_i$ denoting the document at position $i$, the probability of the ranking $R$ then becomes:
\begin{align}
P(R|D) &= \prod^k_{i=1} P(R_i | D \setminus \{ R_1, \ldots, R_{i-1} \}).
\end{align}
After the ranking $R$ has been displayed to the user, they have the option to interact with it.
The user may choose to click on some (or none) of the documents.
Based on these clicks, \ac{\OurMethod} will infer preferences between the displayed documents.
We assume that clicked documents are preferred over observed unclicked documents.
However, to the algorithm it is unknown which unclicked documents the user has considered.
As a solution, \ac{\OurMethod} relies on the assumption that every document preceding a clicked document and the first subsequent unclicked document was observed, as illustrated in Figure~\ref{fig:clickselection}.
This preference assumption has been proven useful in \ac{IR} before, for instance in pairwise \ac{LTR} on click logs \cite{Joachims2002} and recently in online evaluation \cite{oosterhuis2017sensitive}.
We will denote preferences between documents inferred from clicks as: $d_k >_\mathbf{c} d_l$ where $d_k$ is preferred over $d_l$.

Then $\theta$ is updated by optimizing pairwise probabilities over the preference pairs;
for each inferred document preference $d_k >_\mathbf{c} d_l$, the probability that the preferred document $d_k$ is sampled before $d_l$ is sampled is increased \citep{szorenyi2015online}:
\begin{align}
P(d_k \succ d_k)
=
\frac{P(d_k | D)}{P(d_k | D) + P(d_l | D)}
=
\frac{ e^{f(\mathbf{d}_k)} }{
e^{f(\mathbf{d}_k)}
+
e^{f(\mathbf{d}_l)}
}.
\label{eq:pairprob}
\end{align}
We have chosen for pairwise optimization over listwise optimization because a pairwise method can be made unbiased by reweighing preference pairs.
To do this we introduce the weighting function $\rho(d_k, d_l, R, D)$ and estimate the gradient of the user preferences by the weighted sum:
\begin{align}
\begin{split}
&\mbox{}\!\!\! \nabla f_\theta (\cdot)\\
&\approx\!\!
\sum_{ d_k >_\mathbf{c} d_l } \!\!\rho(d_k, d_l, R, D) \left[\nabla P(d_k \succ d_l)\right] \\ 
& =\!\!
\sum_{ d_k >_\mathbf{c} d_l } \!\!\rho(d_k, d_l, R, D)
\frac{ 
e^{f_{\theta}(\mathbf{d}_k)}e^{f_{\theta}(\mathbf{d}_l)}
}{
(e^{f_{\theta}(\mathbf{d}_k)} + e^{f_{\theta}(\mathbf{d}_l)}) ^ 2
}
\!\left(f'_{\theta}(\mathbf{d}_k) - f'_{\theta}(\mathbf{d}_l)\right)
\!.\!\!
\end{split}
\label{eq:novelgradient}
\end{align}
The $\rho$ function is based on the reversed pair ranking $R^*(d_k, d_l, R)$, which is the same ranking as $R$ with the position of $d_k$ and $d_l$ swapped.
An example of a reversed pair ranking is illustrated in Figure~\ref{fig:reversepair}.
The idea is that if a preference for $d_k >_\mathbf{c} d_l$ is inferred in $R$ and both documents are equally relevant, then the reverse preference $d_l >_\mathbf{c} d_k$ is equally likely to be inferred in $R^*(d_k, d_l, R)$.
The $\rho$ function reweighs the found preferences to the ratio between the probabilities of $R$ or $R^*(d_k, d_l, R)$ occurring:
\begin{align}
\rho(d_k, d_l, R, D) &= \frac{P(R^*(d_k, d_l, R)|D)}{P(R|D) + P(R^*(d_k, d_l, R)|D)}.
\label{eq:rho}
\end{align}
This procedure has similarities with importance sampling~\cite{owen2013monte}; 
however, we found that reweighing according to the ratio between $R$ and $R^*$ provides a more stable performance, since it produces less extreme values. 
Section~\ref{sec:method:unbiased} details exactly how $\rho$ creates an unbiased gradient.

\begin{algorithm}[t]
\caption{\acf{\OurMethod}.} 
\label{alg:novel}
\begin{algorithmic}[1]
\STATE \textbf{Input}: initial weights: $\mathbf{\theta}_1$; scoring function: $f$; learning rate $\eta$.  \label{line:novel:initmodel}
\FOR{$t \leftarrow  1 \ldots \infty$ }
	\STATE $q_t \leftarrow \mathit{receive\_query}(t)$\hfill \textit{\small // obtain a query from a user} \label{line:novel:query}
	\STATE $D_t \leftarrow \mathit{preselect\_documents}(q_t)$\hfill \textit{\small // preselect documents for query} \label{line:novel:preselect}
	\STATE $\mathbf{R}_t \leftarrow \mathit{sample\_list}(f_{\theta_t}, D_t)$ \hfill \textit{\small // sample list according to Eq.~\ref{eq:docprob}} \label{line:novel:samplelist}
	\STATE $\mathbf{c}_t \leftarrow \mathit{receive\_clicks}(\mathbf{R}_t)$ \hfill \textit{\small // show result list to the user} \label{line:novel:clicks}
	\STATE $\nabla f_{\theta_{t}} \leftarrow \mathbf{0}$ \hfill \textit{\small // initialize gradient} \label{line:novel:initgrad}
	\FOR{$d_k >_{\mathbf{c}} d_l \in \mathbf{c}_t$} \label{line:novel:prefinfer}
	\STATE $w \leftarrow \rho(d_k, d_l, R, D)$  \hfill \textit{\small // initialize pair weight (Eq.~\ref{eq:rho})} \label{line:novel:initpair}
	\STATE $w \leftarrow w 
            \frac{ 
            e^{f_{\theta_t}(\mathbf{d}_k)}e^{f_{\theta_t}(\mathbf{d}_l)}
            }{
            (e^{f_{\theta_t}(\mathbf{d}_k)} + e^{f_{\theta_t}(\mathbf{d}_l)}) ^ 2
            }$
             \hfill \textit{\small // pair gradient (Eq.~\ref{eq:novelgradient})} \label{line:novel:pairgrad}
	\STATE  $\nabla f_{\theta_{t}} \leftarrow \nabla \theta_t + w (f'_{\theta_t}(\mathbf{d}_k) - f'_{\theta_t}(\mathbf{d}_l))$
	  \hfill \textit{\small // model gradient (Eq.~\ref{eq:novelgradient})} \label{line:novel:modelgrad}
	\ENDFOR
	\STATE $\theta_{t+1} \leftarrow \theta_{t} + \eta \nabla f_{\theta_{t}}$
	\hfill \textit{\small // update the ranking model} \label{line:novel:update}
\ENDFOR
\end{algorithmic}
\end{algorithm}

Algorithm~\ref{alg:novel} describes the \ac{\OurMethod} method step by step:
Given the initial parameters $\theta_1$ and a differentiable scoring function $f$ (Line~\ref{line:novel:initmodel}), the method waits for a user-issued query $q_t$ to arrive (Line~\ref{line:novel:query}).
Then the preselected set of documents $D_t$ for the query is fetched (Line~\ref{line:novel:preselect}), in our experiments these preselections are given in the \ac{LTR} datasets that we use.
A result list $R$ is sampled from the current model (Line~\ref{line:novel:samplelist} and Equation~\ref{eq:docprob}) and displayed to the user.
The clicks from the user are logged (Line~\ref{line:novel:clicks}) and preferences between the displayed documents inferred (Line~\ref{line:novel:prefinfer}).
The gradient is initialized (Line~\ref{line:novel:initgrad}), and for each pair document pair $d_k$, $d_l$ such that $d_k >_{\mathbf{c}} d_l$, the weight $\rho(d_k, d_l, R, D)$ is calculated (Line~\ref{line:novel:initpair} and Equation~\ref{eq:rho}), followed by the gradient for the pair probability (Line~\ref{line:novel:pairgrad} and Equation~\ref{eq:novelgradient}).
Finally, the gradient for the scoring function $f$ is weighted and added to the gradient (Line~\ref{line:novel:modelgrad}), resulting in the estimated gradient.
The model is then updated by taking an $\eta$ step in the direction of the gradient (Line~\ref{line:novel:update}).
The algorithm again waits for the next query to arrive and thus the process continues indefinitely.

\ac{\OurMethod} has some notable advantages over \acf{MGD}~\citep{schuth2016mgd}.
Firstly, it explicitly models uncertainty over the documents per query, thus \ac{\OurMethod} is able to have high confidence in its ranking for one query, while being completely uncertain for another query.
As a result, it will vary the amount of exploration per query, allowing it to avoid exploration in cases where it is not required and focussing on areas where it can improve.
In contrast, \ac{MGD} does not explicitly model confidence: its degree of exploration is only affected by the norm of its linear model \citep{oosterhuis2017balancing}.
Consequently, \ac{MGD} is unable to vary exploration per query nor is there a way to directly measure its level of confidence.
Secondly, \ac{\OurMethod} works for any differentiable scoring function $f$ and does not rely on sampling model variants.
Conversely, \ac{MGD} is based around sampling from the unit sphere around a model; this approach is very ineffective for non-linear models.
Additionally, sampling large models and producing rankings for them can be very computationally expensive.
Besides these beneficial properties, our experimental results in Section~\ref{sec:results} show that \ac{\OurMethod} achieves significantly higher levels of performance than \ac{MGD} and other previous methods.

\subsection{Unbiased gradient estimation}
\label{sec:method:unbiased}

The previous section introduced \ac{\OurMethod}; this section answers \ref{rq:unbiased}: is the gradient estimation of \ac{\OurMethod} unbiased?

First, Theorem~\ref{theorem:unbiased} will provide a definition of unbiasedness w.r.t.\ user document pair preferences.
Then we state the assumptions we make about user behavior and use them to prove Theorem~\ref{theorem:unbiased}.

\begin{theorem}
The expected estimated gradient of \ac{\OurMethod} can be written as a weighted sum, with a unique weight $\alpha_{k,l}$ for each possible document pair $d_k$ and $d_l$ in the document collection $D$:
\begin{align}
E[\nabla f_\theta(\cdot)] = \sum_{d_k, d_l \in D} \alpha_{k,l} (f'_{\theta_t}(\mathbf{d}_k) - f'_{\theta_t}(\mathbf{d}_l)).
\label{eq:theorem:pair}
\end{align}
The signs of the weights $\alpha_{k,l}$  adhere to user preferences between documents. 
That is, if there is no preference:
\begin{align}
d_k =_{rel} d_l \Leftrightarrow \alpha_{k,l} = 0;
\label{eq:theorem:equal}
\end{align}
if $d_k$ is preferred over $d_l$:
\begin{align}
d_k >_{rel} d_l \Leftrightarrow \alpha_{k,l} > 0; 
\label{eq:theorem:greater}
\end{align}
and if $d_l$ is preferred over $d_k$:
\begin{align}
d_k <_{rel} d_l \Leftrightarrow \alpha_{k,l} < 0.
\label{eq:theorem:lesser}
\end{align}
Therefore, in expectation \ac{\OurMethod} will perform updates that adhere to the preferences between the documents in every possible document pair.
\label{theorem:unbiased}
\end{theorem}

\paragraph{Assumptions.}
To prove Theorem~\ref{theorem:unbiased} the following assumptions about user behavior will be used:

\smallskip
\subparagraph{Assumption 1.} 
We assume that clicks from a user are position biased and conditioned on the relevance of the current document and the previously considered documents.
For a click on a document in ranking $R$ at position $i$ the probability can be written as:
\begin{align}
P(\textit{click}(R_i) | \{R_0, \ldots, R_{i-1}, R_{i+1}\}).
\label{eq:userassumption}
\end{align}
For ease of notation, we will denote the set of ``other documents'' as $\OtherDocuments{}$ from here on.

\subparagraph{Assumption 2.} 
If there is no user preference between two documents $d_k, d_l$, denoted by $d_k =_\textit{rel} d_l$, we assume that each is equally likely to be clicked given the same context:
\begin{align}
d_k =_\textit{rel} d_l \Rightarrow P(\textit{click}(d_k)| \OtherDocuments{}) = P(\textit{click}(d_l) | \OtherDocuments{} ).
\label{eq:equalsame}
\end{align}

\subparagraph{Assumption 3.}
If a document in the set of documents being considered is replaced with an equally preferred document the click probability is not affected:
\begin{align}
d_k =_\textit{rel} d_l \Rightarrow  P(\textit{click}(R_i)| \{\ldots, d_k\} ) &= P(\textit{click}(R_i)|  \{\ldots, d_l\} ).
\label{eq:equaladded}
\end{align}

\subparagraph{Assumption 4.}
Similarly, given the same context if one document is preferred over another, then it is more likely to be clicked:
\begin{align}
d_k >_\textit{rel} d_l \Rightarrow P(\textit{click}(d_k)| \OtherDocuments{}) > P(\textit{click}(d_l)| \OtherDocuments{} ).
\label{eq:greatersame}
\end{align}

\subparagraph{Assumption 5.}
Lastly, for any pair $d_k >_\textit{rel} d_l$, the considered document set $\{\ldots, d_k\}$ and the same set with $d_k$ replaced $\{\ldots, d_l\}$.
We assume that the preferred $d_k$ in the context of $\{\ldots, d_l\}$ is more likely to be clicked than $d_l$ in the context of $\{\ldots, d_k\}$:
\begin{align}
d_k >_\textit{rel} d_l \Rightarrow  P(\textit{click}(d_k)| \{\ldots, d_k\} ) > P(\textit{click}(d_l)|  \{\ldots, d_l\} ).
\label{eq:greateradded}
\end{align}
These are all the assumptions we make about the user. 
With these assumptions, we can proceed to prove Theorem~\ref{theorem:unbiased}.

\begin{proof}[Proof of Theorem~\ref{theorem:unbiased}.]
We denote the probability of inferring the preference of $d_k$ over $d_l$ in ranking $R$ as $P(d_k >_\mathbf{c} d_l | R)$. 
Then the expected gradient $\nabla f_\theta(\cdot)$ of \ac{\OurMethod} can be written as:
\begin{align}
\begin{split}
E[\nabla f_\theta(\cdot)] = \sum_R & \sum_{d_k, d_l \in D} \left[ \vphantom{\frac{\delta}{\delta \theta}} \right. 
P(d_k >_\mathbf{c} d_l | R) \cdot P(R) \cdot {} \\
& \mbox{}\hspace*{0.2cm} \rho(d_k, d_l, R, D) \cdot 
 \left.\vphantom{\sum_R}\left[\nabla P(d_k \succ d_l)\right]\right] .
\end{split}
\end{align}
We will rewrite this expectation using the symmetry property of the reversed pair ranking:
\begin{align}
R^n = R^*(d_k, d_l, R^m) \Leftrightarrow R^m = R^*(d_k, d_l, R^n).
\end{align}
First, we define a weight $\omega_{k,l}^R$ for every document pair $d_k,d_l$ and ranking $R$ so that:
\begin{align}
\begin{split}
\omega_{k,l}^R & = P(R) \rho(d_k, d_l, R, D) \\
&= \frac{P(R|D)P(R^*(d_k, d_l, R)|D)}{P(R|D) + P(R^*(d_k, d_l, R)|D)}.
\end{split}
\end{align}
Therefore, the weight for the reversed pair ranking is equal:
\begin{align}
\begin{split}
\omega_{k,l}^{R^*(d_k, d_l, R)} &= P(R^*(d_k, d_l, R)) \rho(d_k, d_l, R^*(d_k, d_l, R), D) \\
&= \omega_{k,l}^R.
\end{split}
\end{align}
Then, using the symmetry of Equation~\ref{eq:pairprob} we see that:
\begin{align}
\nabla P(d_k \succ d_l) = -\nabla P(d_l \succ  d_k).
\end{align}
Thus, with $R^*$ as a shorthand for $R^*(d_k, d_l, R)$, the expectation can be rewritten as:
\begin{align}
\begin{split}
& E[\nabla f_\theta(\cdot)] = \\
& \qquad \sum_{d_k, d_l \in D} \sum_R \frac{\omega_{i,j}^R}{2}  
  \left( \vphantom{\frac{\omega_{i,j}^R}{2}  }P(d_k >_\mathbf{c} d_l | R)  - P(d_l >_\mathbf{c} d_k | R^*)\right) \cdot{}  \\
&\mbox{}\hspace*{5.4cm} \left[\vphantom{\frac{\omega_{i,j}^R}{2}  }\nabla P(d_k \succ d_l)\right],
\end{split}
\end{align}
proving that the expected gradient matches the form of Equation~\ref{eq:theorem:pair}.
Then to prove that Equations~\ref{eq:theorem:equal},~\ref{eq:theorem:greater}, and~\ref{eq:theorem:lesser} are correct we will show that:
\begin{align}
d_k =_\textit{rel} d_l &\Rightarrow  P(d_k >_\mathbf{c} d_l | R) = P(d_l >_\mathbf{c} d_k | R^*),
\label{eq:equalrequirement}
\\
d_k >_\textit{rel} d_l &\Rightarrow  P(d_k >_\mathbf{c} d_l | R) > P(d_l >_\mathbf{c} d_k | R^*),
\label{eq:greaterrequirement}
\\
d_k <_\textit{rel} d_l &\Rightarrow  P(d_k >_\mathbf{c} d_l | R) < P(d_l >_\mathbf{c} d_k | R^*).
\label{eq:lesserrequirement}
\end{align}
If a preference $R_i >_\mathbf{c} R_j$ is inferred then there are only three possible cases based on the positions:
\begin{enumerate}
\item The clicked document succeeds the unclicked document by more than one position: $i + 1 > j$.
\item The clicked document precedes the unclicked document by more than one position: $i - 1 < j$.
\item The clicked document is one position before or after the unclicked document: $i = j +1 \lor i = j-1$.
\end{enumerate}
In the first case the clicked document succeeds the other by more than one position, the probability of an inferred preference is then:
\begin{align}
\begin{split}
i + 1 > j 
&\Rightarrow
P(R_i >_\mathbf{c} R_j | R) = {}\\
&\qquad P(\mathbf{c}_i | R_i, \{\ldots, R_j\}\ )\cdot{}\\
&\qquad (1- P(\mathbf{c}_j | R_j, \OtherDocuments{} )).
\end{split}
\label{eq:simplecase}
\end{align}
Combining Assumption~2~and~3 with Equation~\ref{eq:simplecase} proves Equation~\ref{eq:equalrequirement} for this case.
Furthermore, combining Assumption~4~and~5 with Equation~\ref{eq:simplecase} proves Equations~\ref{eq:greaterrequirement}~and~\ref{eq:lesserrequirement} for this case as well.

Then the second case is when the clicked document appears more than one position before the unclicked document, the probability of the inferred preference is then:
\begin{align}
\begin{split}
i + 1 < j & \Rightarrow P(R_i >_\mathbf{c} R_j | R) = {}\\
&\qquad P(\mathbf{c}_i | R_i, \OtherDocuments{} ) \cdot{} \\
&\qquad (1- P(\mathbf{c}_j | R_j,\{\ldots, R_i\}\ ))\cdot {} \\
&\qquad P(\mathbf{c}_\textit{rem}),
\end{split}
\label{eq:complexcase}
\end{align}
where $P(\mathbf{c}_\textit{rem})$ denotes the probability of an additional click that is required to add $R_j$ to the inferred observed documents.
First, due to Assumption~1 this probability will be the same for $R$ and $R^*$:
\begin{align}
P(\mathbf{c}_\textit{rem}| R_i, R_j , R) = P(\mathbf{c}_\textit{rem}| R_i, R_j , R^*).
\end{align}
Combining Assumption~2~and~3 with Equation~\ref{eq:complexcase} also proves Equation~\ref{eq:equalrequirement} for this case.
Furthermore, combining Assumption~4~and~5 with Equation~\ref{eq:complexcase} also proves Equation~\ref{eq:greaterrequirement}~and~\ref{eq:lesserrequirement} for this case as well.

Lastly, in the third case the clicked document is one position before or after the other document, the probability of the inferred preference is then:
\begin{align}
\begin{split}
i = j +1 \lor i = j-1 
&\Rightarrow
P(R_i >_\mathbf{c} R_j | R) = {}\\
&\qquad P(\mathbf{c}_i | R_i, \{\ldots, R_j\}\ )\cdot{}\\
&\qquad (1- P(\mathbf{c}_j | R_j, \{\ldots, R_i\} )).
\end{split}
\label{eq:specialcase}
\end{align}
Combining Assumption~3 with Equation~\ref{eq:specialcase} proves Equation~\ref{eq:equalrequirement} for this case as well.
Then, combining Assumption~5 with Equation~\ref{eq:specialcase} also proves Equation~\ref{eq:greaterrequirement}~and~\ref{eq:lesserrequirement} for this case.
\end{proof}


\noindent%
This concludes our proof of the unbiasedness of \ac{\OurMethod}. 
Hence, we answer \ref{rq:unbiased} positively: the gradient estimation of \ac{\OurMethod} is unbiased.
We have shown that the expected gradient is in line with user preferences between document pairs.

\begin{table}[tb]
\caption{Instantiations of Cascading Click Models~\citep{guo09:efficient} as used for simulating user behavior in experiments.}
\centering
\begin{tabularx}{\columnwidth}{ X c c c c c c c c c c }
\toprule
& \multicolumn{5}{c}{ $P(\mathit{click}=1\mid R)$} & \multicolumn{5}{c}{ $P(\mathit{stop}=1\mid click=1,  R)$} \\
\cmidrule(lr){2-6} \cmidrule(l){7-11}
$R$ & \emph{$ 0$} & \emph{$ 1$}  &  \emph{$ 2$} & \emph{$ 3$} & \emph{$ 4$}
 & \emph{$0$} & \emph{$ 1$} & \emph{$ 2$} & \emph{$ 3$} & \emph{$ 4$} \\
\midrule
 \emph{perf} &  0.0 &  0.2 &  0.4 &  0.8 &  1.0 &  0.0 &  0.0 &  0.0 &  0.0 &  0.0 \\
 \emph{nav} &  \phantom{5}0.05 &  0.3 &  0.5 &  0.7 &  \phantom{5}0.95 &  0.2 &  0.3 &  0.5 &  0.7 &  0.9 \\
 \emph{inf} &  0.4 &  0.6 &  0.7 &  0.8 &  0.9 &  0.1 &  0.2 &  0.3 &  0.4 &  0.5 \\
\bottomrule
\end{tabularx}
\label{tab:clickmodels}
\end{table}

\section{Experiments}
\label{sec:experiments}

In this section we detail the experiments that were performed to answer the research questions in Section~\ref{sec:intro}.

\subsection{Datasets}
\label{sec:experiments:datasets}

Our experiments are performed over five publicly available \acs{LTR} datasets; we have selected three large labelling sets from commercial search engines and two smaller research datasets.
Every dataset consists of a set of queries with each query having a corresponding preselected document set.
The exact content of the queries and documents are unknown, each query is represented only by an identifier, but each query-document pair has a feature representation and relevance label.
Depending on the dataset, the relevance labels are graded differently; we have purposefully chosen datasets that have at least two grades of relevance.
Each dataset is divided in training, validation and test partitions.

The oldest datasets we use are \emph{MQ2007} and \emph{MQ2008}~\cite{qin2013introducing} which are based on the Million Query Track \cite{allan2007million} and consist of \numprint{1700} and 800 queries. They use representations of 46~features that encode ranking models such as TF.IDF, BM25, Language Modeling, Page\-Rank, and HITS on different parts of the documents. They are divided into five folds and the labels are on a three-grade scale from \emph{not relevant} (0) to \emph{very relevant}~(2).

In 2010 Microsoft released the \emph{MSLR-WEB30k} and \emph{MLSR-WEB10K} datasets \cite{qin2013introducing}, which are both created from a retired labelling set of a commercial web search engine (Bing).
The former contains \numprint{30000} queries with each query having 125~assessed documents on average, query-document pairs are encoded in 136~features,
The latter is a subsampling of \numprint{10000} queries from the former dataset.
For practical reasons only \emph{MLSR-WEB10K} was used for this paper.
Also in 2010 Yahoo!\ released an \ac{LTR} dataset \cite{Chapelle2011}.
It consists of \numprint{29921} queries and \numprint{709877} documents encoded in 700~features, all sampled from query logs of the Yahoo! search engine.
Finally, in 2016 a \ac{LTR} dataset released by the Istella search engine \cite{dato2016fast}.
It is the largest with \numprint{33118} queries, an average of~315 documents per query and 220~features.
These three commercial datasets all label relevance on a five-grade scale: from \emph{not relevant} (0) to \emph{perfectly relevant}~(4).

\subsection{Simulating user behavior}
\label{sec:experiments:users}

For simulating users we follow the standard setup for \acs{OLTR} simulations \cite{He2009,hofmann11:balancing,schuth2016mgd,oosterhuis2016probabilistic,zoghi:wsdm14:relative}.
First, queries issued by  users are simulated by uniformly sampling from the static dataset.
Then the algorithm determines the result list of documents to display.
User interactions with the displayed list are then simulated using a \emph{cascade click model}~\cite{chuklin-click-2015,guo09:efficient}.
This models a user who goes through the documents one at a time in the displayed order.
At each document, the user decides whether to click it or not, modelled as a probability conditioned on the relevance label $R$: $P(click=1\mid R)$.
After a click has occurred, the user's information need may be satisfied and they may then stop considering documents.
The probability of a user stopping after a click is modelled as $P(stop=1\mid click=1,  R)$.
For our experiments $\kappa=10$ documents are displayed at each impression.

The three instantiations of cascade click models that we used are listed in Table~\ref{tab:clickmodels}.
First, a \emph{perfect} user is modelled who considers every document and solely clicks on all relevant documents.
The second models a user with a \emph{navigational} task, where a single highly relevant document is searched.
Finally, an \emph{informational} instantiation models a user without a specific information need, and thus typically clicks on many documents.
These models have varying levels of noise, as each behavior depends on the relevance labels of documents with a different degree.

\begin{figure}[tb]
\centering
\includegraphics[scale=0.345]{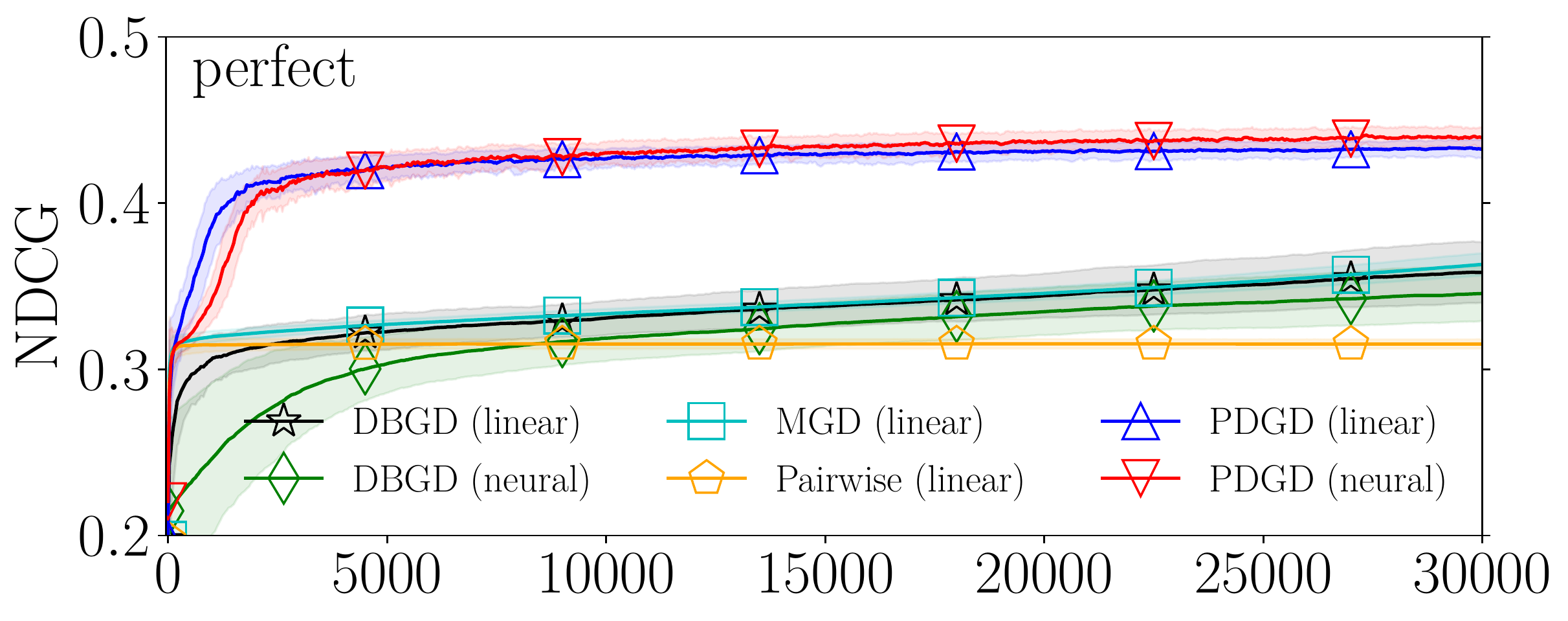}
\includegraphics[scale=0.345]{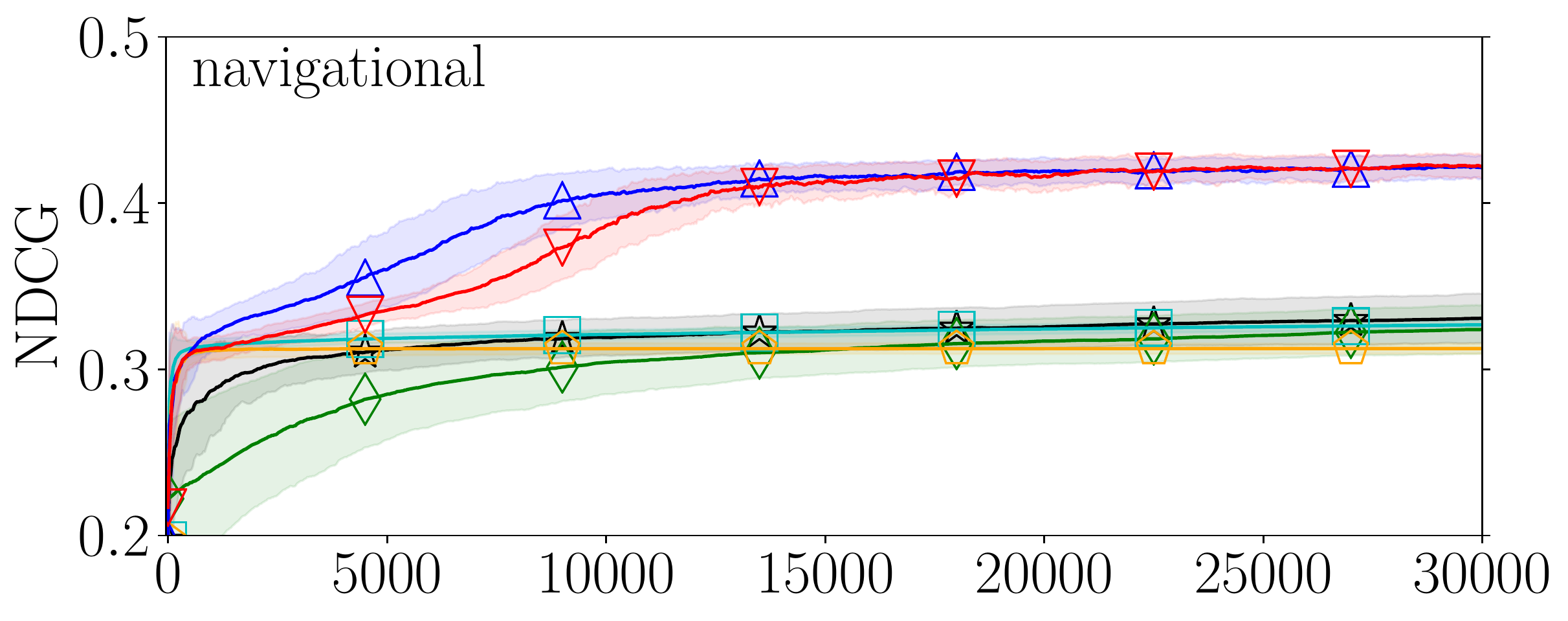}
\includegraphics[scale=0.345]{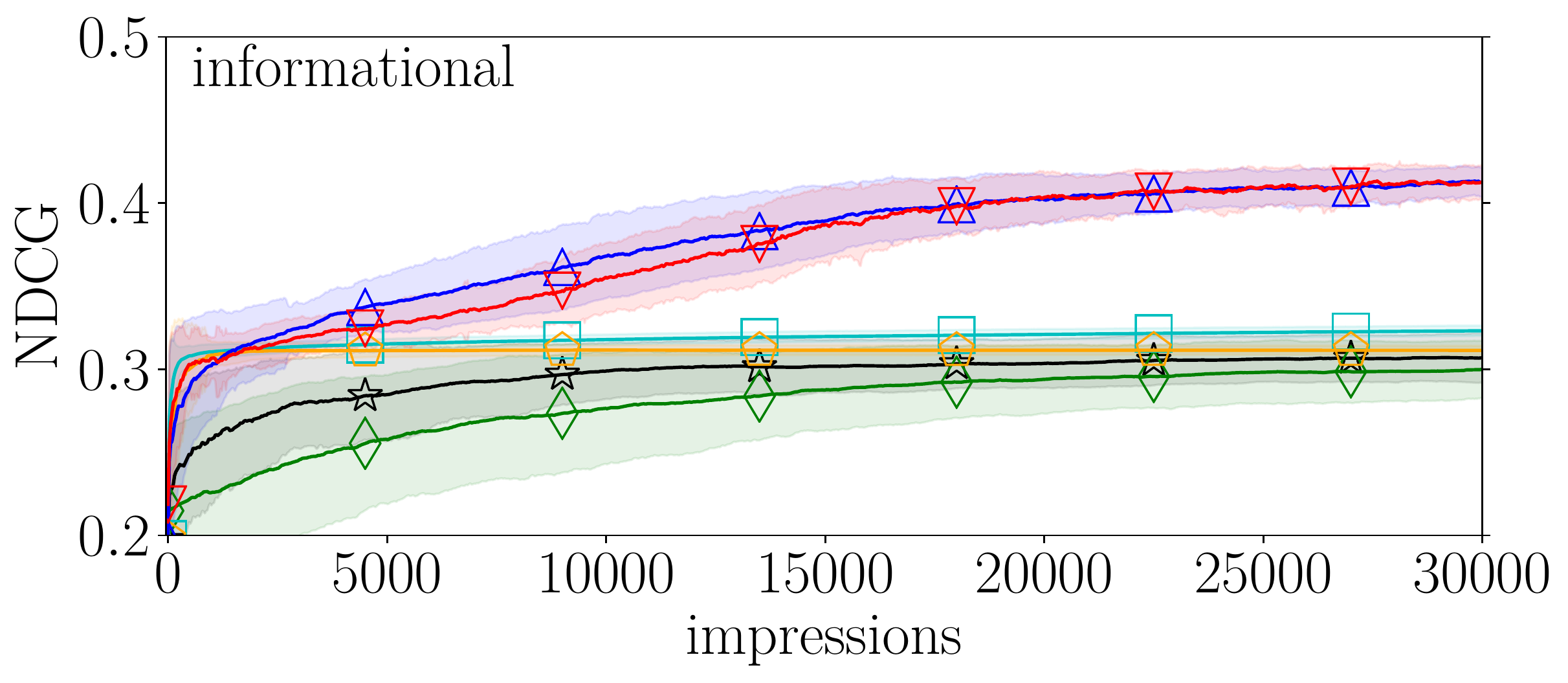}
\caption{Offline performance (NDCG) on the MSLR-WEB10k dataset under three different click models, the shaded areas indicate the standard deviation.}
\label{fig:offline}
\end{figure}

\begin{table*}[t]
\centering
\caption{Offline performance (NDCG) for different instantiations of CCM (Table~\ref{tab:clickmodels}). The standard deviation is shown in brackets, bold values indicate the highest performance per dataset and click model, significant improvements over the \acs{DBGD}, \acs{MGD} and pairwise baselines are indicated by  \enkelop\ (p $<$ 0.05) and \dubbelop\ (p $<$ 0.01), no losses were measured.}
\begin{tabular*}{\textwidth}{@{\extracolsep{\fill} } l  l l l l l  }
\toprule
 & { \small \textbf{MQ2007}}  & { \small \textbf{MQ2008}}  & { \small \textbf{MSLR-WEB10k}}  & { \small \textbf{Yahoo}}  & { \small \textbf{istella}} \\
\midrule
& \multicolumn{5}{c}{\textit{perfect}} \\
\midrule
DBGD (linear) & 0.483 {\tiny (0.023)} & 0.683 {\tiny (0.024)} & 0.331 {\tiny (0.010)} & 0.684 {\tiny (0.010)} & 0.448 {\tiny (0.014)} \\
DBGD (neural) & 0.463 {\tiny (0.025)} & 0.670 {\tiny (0.026)} & 0.319 {\tiny (0.014)} & 0.676 {\tiny (0.016)} & 0.429 {\tiny (0.017)} \\
MGD (linear) & 0.494 {\tiny (0.022)} & 0.690 {\tiny (0.019)} & 0.333 {\tiny (0.003)} & 0.714 {\tiny (0.002)} & 0.496 {\tiny (0.004)} \\
Pairwise (linear) & 0.479 {\tiny (0.022)} & 0.674 {\tiny (0.017)} & 0.315 {\tiny (0.003)} & 0.709 {\tiny (0.001)} & 0.252 {\tiny (0.002)} \\
PDGD (linear) & \bf 0.511 {\tiny (0.017)} {\tiny \dubbelop} {\tiny \dubbelop} {\tiny \dubbelop} {\tiny \dubbelop} & \bf 0.699 {\tiny (0.024)} {\tiny \dubbelop} {\tiny \dubbelop} {\tiny \dubbelop} {\tiny \dubbelop} & 0.427 {\tiny (0.005)} {\tiny \dubbelop} {\tiny \dubbelop} {\tiny \dubbelop} {\tiny \dubbelop} & \bf 0.736 {\tiny (0.004)} {\tiny \dubbelop} {\tiny \dubbelop} {\tiny \dubbelop} {\tiny \dubbelop} & 0.573 {\tiny (0.004)} {\tiny \dubbelop} {\tiny \dubbelop} {\tiny \dubbelop} {\tiny \dubbelop} \\
PDGD (neural) & 0.509 {\tiny (0.020)} {\tiny \dubbelop} {\tiny \dubbelop} {\tiny \dubbelop} {\tiny \dubbelop} & 0.698 {\tiny (0.024)} {\tiny \dubbelop} {\tiny \dubbelop} {\tiny \dubbelop} {\tiny \dubbelop} & \bf 0.430 {\tiny (0.006)} {\tiny \dubbelop} {\tiny \dubbelop} {\tiny \dubbelop} {\tiny \dubbelop} & 0.733 {\tiny (0.005)} {\tiny \dubbelop} {\tiny \dubbelop} {\tiny \dubbelop} {\tiny \dubbelop} & \bf 0.575 {\tiny (0.006)} {\tiny \dubbelop} {\tiny \dubbelop} {\tiny \dubbelop} {\tiny \dubbelop} \\
\midrule
& \multicolumn{5}{c}{\textit{navigational}} \\
\midrule
DBGD (linear) & 0.461 {\tiny (0.025)} & 0.670 {\tiny (0.025)} & 0.319 {\tiny (0.011)} & 0.661 {\tiny (0.023)} & 0.401 {\tiny (0.015)} \\
DBGD (neural) & 0.430 {\tiny (0.033)} & 0.646 {\tiny (0.031)} & 0.304 {\tiny (0.019)} & 0.649 {\tiny (0.029)} & 0.382 {\tiny (0.024)} \\
MGD (linear) & 0.426 {\tiny (0.020)} & 0.662 {\tiny (0.015)} & 0.321 {\tiny (0.003)} & 0.706 {\tiny (0.009)} & 0.405 {\tiny (0.004)} \\
Pairwise (linear) & 0.476 {\tiny (0.022)} & 0.677 {\tiny (0.018)} & 0.312 {\tiny (0.003)} & 0.696 {\tiny (0.004)} & 0.209 {\tiny (0.002)} \\
PDGD (linear) & \bf 0.496 {\tiny (0.019)} {\tiny \dubbelop} {\tiny \dubbelop} {\tiny \dubbelop} {\tiny \dubbelop} & \bf 0.695 {\tiny (0.021)} {\tiny \dubbelop} {\tiny \dubbelop} {\tiny \dubbelop} {\tiny \dubbelop} & \bf 0.406 {\tiny (0.015)} {\tiny \dubbelop} {\tiny \dubbelop} {\tiny \dubbelop} {\tiny \dubbelop} & \bf 0.725 {\tiny (0.005)} {\tiny \dubbelop} {\tiny \dubbelop} {\tiny \dubbelop} {\tiny \dubbelop} & \bf 0.540 {\tiny (0.008)} {\tiny \dubbelop} {\tiny \dubbelop} {\tiny \dubbelop} {\tiny \dubbelop} \\
PDGD (neural) & 0.493 {\tiny (0.020)} {\tiny \dubbelop} {\tiny \dubbelop} {\tiny \dubbelop} {\tiny \dubbelop} & 0.692 {\tiny (0.019)} {\tiny \dubbelop} {\tiny \dubbelop} {\tiny \dubbelop} {\tiny \dubbelop} & 0.386 {\tiny (0.019)} {\tiny \dubbelop} {\tiny \dubbelop} {\tiny \dubbelop} {\tiny \dubbelop} & 0.722 {\tiny (0.006)} {\tiny \dubbelop} {\tiny \dubbelop} {\tiny \dubbelop} {\tiny \dubbelop} & 0.532 {\tiny (0.011)} {\tiny \dubbelop} {\tiny \dubbelop} {\tiny \dubbelop} {\tiny \dubbelop} \\
\midrule
& \multicolumn{5}{c}{\textit{informational}} \\
\midrule
DBGD (linear) & 0.411 {\tiny (0.036)} & 0.631 {\tiny (0.036)} & 0.299 {\tiny (0.017)} & 0.620 {\tiny (0.035)} & 0.360 {\tiny (0.028)} \\
DBGD (neural) & 0.383 {\tiny (0.047)} & 0.595 {\tiny (0.053)} & 0.276 {\tiny (0.033)} & 0.603 {\tiny (0.040)} & 0.316 {\tiny (0.057)} \\
MGD (linear) & 0.406 {\tiny (0.021)} & 0.647 {\tiny (0.036)} & 0.318 {\tiny (0.003)} & 0.676 {\tiny (0.043)} & 0.387 {\tiny (0.005)} \\
Pairwise (linear) & 0.478 {\tiny (0.022)} & 0.677 {\tiny (0.018)} & 0.311 {\tiny (0.003)} & 0.690 {\tiny (0.006)} & 0.183 {\tiny (0.001)} \\
PDGD (linear) & \bf 0.487 {\tiny (0.021)} {\tiny \dubbelop} {\tiny \dubbelop} {\tiny \dubbelop} {\tiny \dubbelop} & \bf 0.690 {\tiny (0.022)} {\tiny \dubbelop} {\tiny \dubbelop} {\tiny \dubbelop} {\tiny \dubbelop} & \bf 0.368 {\tiny (0.025)} {\tiny \dubbelop} {\tiny \dubbelop} {\tiny \dubbelop} {\tiny \dubbelop} & \bf 0.713 {\tiny (0.008)} {\tiny \dubbelop} {\tiny \dubbelop} {\tiny \dubbelop} {\tiny \dubbelop} & \bf 0.532 {\tiny (0.010)} {\tiny \dubbelop} {\tiny \dubbelop} {\tiny \dubbelop} {\tiny \dubbelop} \\
PDGD (neural) & 0.483 {\tiny (0.022)} {\tiny \dubbelop} {\tiny \dubbelop} {\tiny \dubbelop} \hphantom{\tiny \dubbelneer} & 0.686 {\tiny (0.022)} {\tiny \dubbelop} {\tiny \dubbelop} {\tiny \dubbelop} {\tiny \dubbelop} & 0.355 {\tiny (0.021)} {\tiny \dubbelop} {\tiny \dubbelop} {\tiny \dubbelop} {\tiny \dubbelop} & 0.709 {\tiny (0.009)} {\tiny \dubbelop} {\tiny \dubbelop} {\tiny \dubbelop} {\tiny \dubbelop} & 0.525 {\tiny (0.012)} {\tiny \dubbelop} {\tiny \dubbelop} {\tiny \dubbelop} {\tiny \dubbelop} \\
\bottomrule
\end{tabular*}

\label{tab:offline}
\end{table*}

\subsection{Experimental runs}
\label{sec:experiments:runs}

For our experiments three baselines are used. 
First, \ac{MGD} with Probabilistic Multileaving~\citep{oosterhuis2016probabilistic}; this is the highest performing existing \ac{OLTR} method~\citep{oosterhuis2016probabilistic, oosterhuis2017balancing}.
For this work $n=49$ candidates were sampled per iteration from the unit sphere with $\delta=1$; updates are performed with $\eta = 0.01$ and zero initialization was used.
Additionally, \ac{DBGD} is used for comparison as it is one of the most influential methods, it was run with the same parameters except that only $n=1$ candidate is sampled per iteration.
Furthermore, we also let \ac{DBGD} optimize a single hidden-layer neural network with 64 hidden nodes and sigmoid activation functions with \emph{Xavier} initialization~\cite{glorot2010understanding}.
These parameters were also found most effective in previous work~\citep{yue09:inter, hofmann11:balancing, schuth2016mgd, oosterhuis2016probabilistic}.

Additionally, the pairwise method introduced by \citet{hofmann11:balancing} is used as a baseline. Despite not showing significant improvements over \ac{DBGD} in the past~\cite{hofmann11:balancing}, the comparison with \ac{\OurMethod} is interesting because they both estimate gradients from pairwise preferences. 
For this baseline, $\eta = 0.01$ and $\epsilon = 0.8$ is used; these parameters are chosen to maximize the performance at convergence~\cite{hofmann11:balancing}.

Runs with \ac{\OurMethod} are performed with both a linear and neural ranking model.
For the linear ranking model $\eta = 0.1$ and zero initialization was used.
The neural network has the same parameters as the one optimized by \ac{DBGD}, except for $\eta = 0.1$.

\subsection{Metrics and tests}
\label{sec:experiments:evaluation}

Two aspects of performance are evaluated seperately: the final convergence and the ranking quality during training.

Final convergence is addressed in \emph{offline performance} which is the average NDCG@10 of the ranking model over the queries in the held-out test-set.
The offline performance is measured after 10,000 impressions at which point most ranking models have reached convergence.
The user experience during optimization should be considered as well, since deterring users during training would compromise the goal of \ac{OLTR}.
To address this aspect of evaluation \emph{online performance} has been introduced~\citep{Hofmann2013a}; it is the cumulative discounted NDCG@10 of the rankings displayed during training.
For $T$ sequential queries with $R^t$ as the ranking displayed to the user at timestep $t$, this is:
\begin{align}
\mathit{Online\_Performance} = \sum_{t=1}^T \mathit{NDCG}(R^t) \cdot \gamma^{(t-1)}.
\end{align}
This metric models the expected reward a user receives with a $\gamma$ probability that the user stops searching after each query.
We follow previous work~\cite{oosterhuis2017balancing, oosterhuis2016probabilistic} by choosing a discount factor of $\gamma = 0.9995$, consequently queries beyond the horizon of \numprint{10000} queries have a less than $1\%$ impact.

Lastly, all experimental runs are repeated 125 times, spread evenly over the available dataset folds.
Results are averaged and a two-tailed Student's t-test is used for significance testing.
In total, our results are based on more than \numprint{90000000} user impressions.

\begin{figure}[tb]
\centering
\includegraphics[scale=0.345]{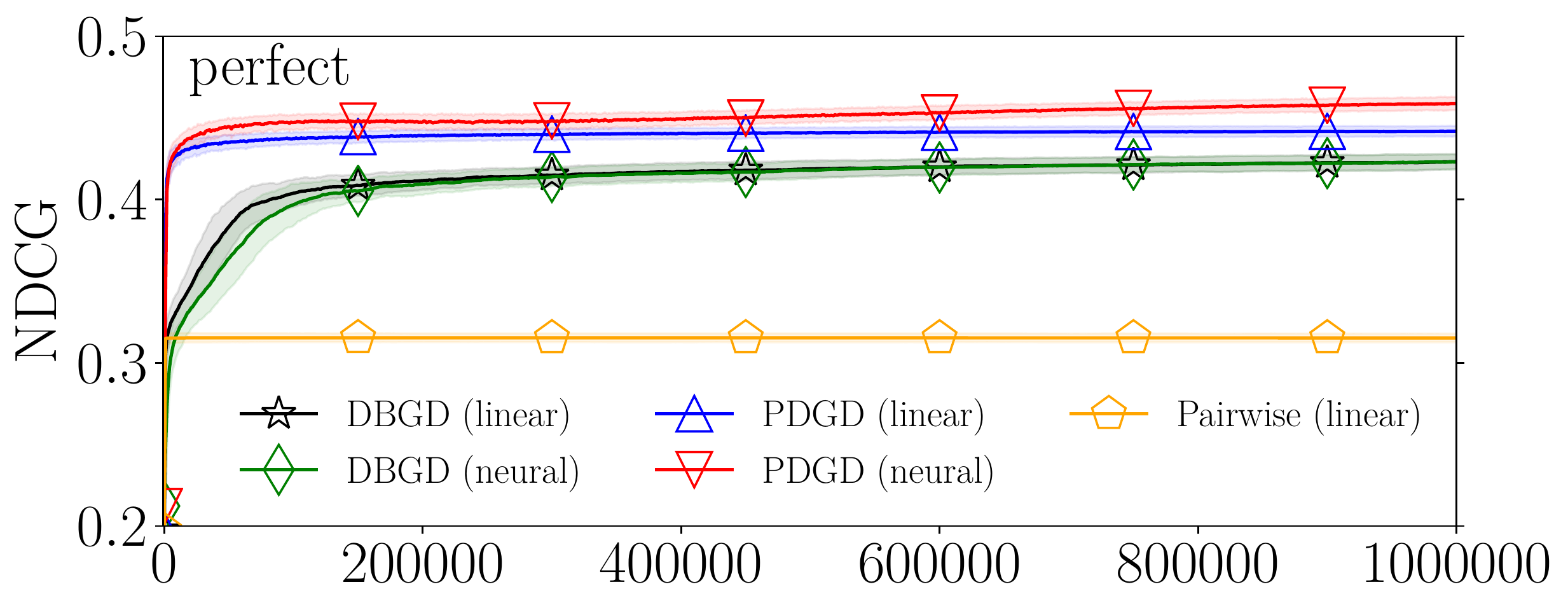}
\includegraphics[scale=0.345]{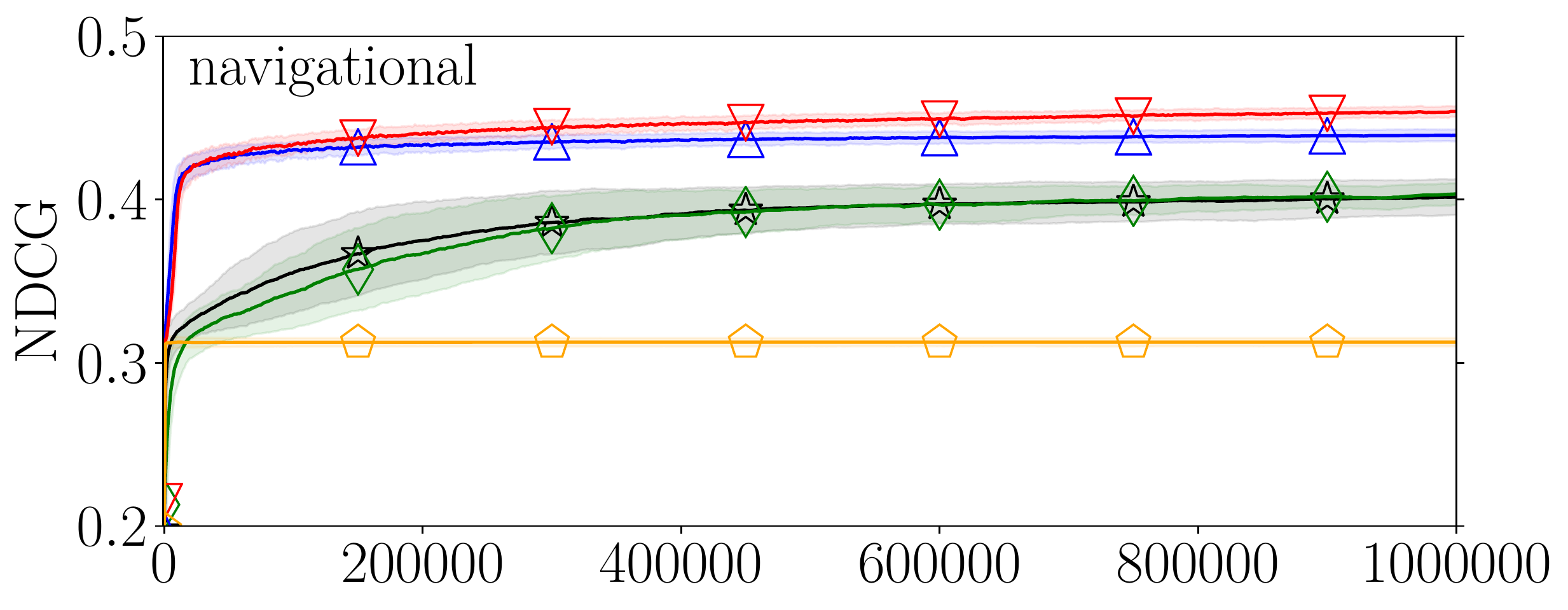}
\includegraphics[scale=0.345]{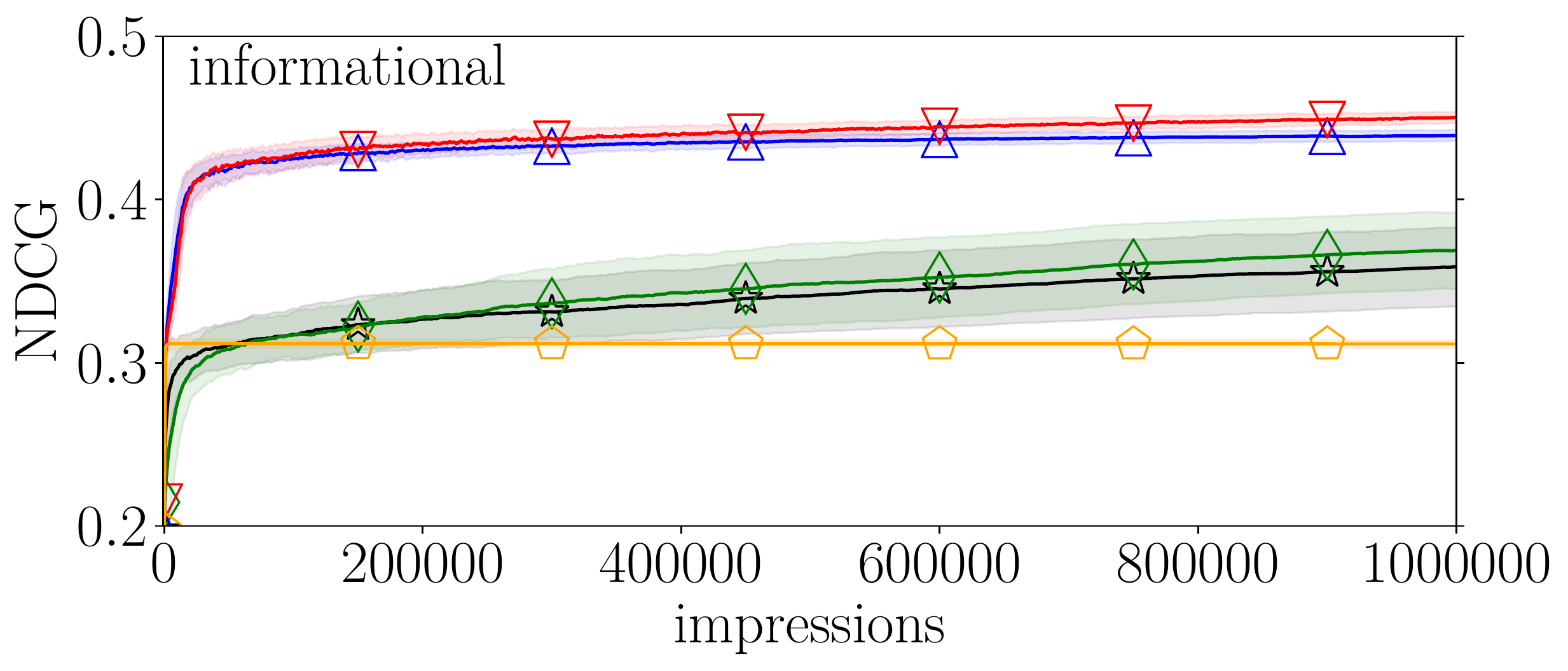}
\caption{Long-term offline performance (NDCG) on the MSLR-WEB10k dataset under three click models, the shaded areas indicate the standard deviation.}
\label{fig:long}
\end{figure}

\section{Results and Analysis}
\label{sec:results}

Our main results are displayed in Table~\ref{tab:offline} and Table~\ref{tab:online}, showing the offline and online performance of all methods, respectively.
Additionally, Figure~\ref{fig:offline} displays the offline performance on the MSLR-WEB10k dataset over \numprint{30000} impressions and Figure~\ref{fig:long} over \numprint{1000000} impressions.
We use these results to answer \ref{rq:performance} -- whether \ac{\OurMethod} provides significant improvements over existing \ac{OLTR} methods -- and \ref{rq:nonlinear} -- whether \ac{\OurMethod} is successful at optimizing different types of ranking models.

\subsection{Convergence of ranking models}
First, we consider the offline performance after \numprint{10000} impressions as reported in Table~\ref{tab:offline}.
We see that the \ac{DBGD} and \ac{MGD} baselines reach similar levels of performance, with marginal differences at low levels of noise.
Our results seem to suggest that \ac{MGD} provides an efficient alternative to \ac{DBGD} that requires fewer user interactions and is more robust to noise.
However, \ac{MGD} does not appear to have an improved point of convergence over \ac{DBGD}, Figure~\ref{fig:offline} further confirms this conclusion.
Additionally, Table~\ref{tab:offline} and Figure~\ref{fig:long} reveal thats \ac{DBGD} is incapable of training its neural network so that it improves over the linear model, even after \numprint{1000000} impressions.

Alternatively, the pairwise baseline displays different behavior, providing improvements over \ac{DBGD} and \ac{MGD} on most datasets under all levels of noise.
However, on the istella dataset large decreases in performance are observed.
Thus it is unclear if this method provides a reliable alternative to  \ac{DBGD} or \ac{MGD} in terms of convergence.
Figure~\ref{fig:offline} also reveals that it converges within several hundred impressions, while  \ac{DBGD} or \ac{MGD} continue to learn and considerably improve over the total  \numprint{30000} impressions.
Because the pairwise baseline also converges sub-optimally under the perfect click model, we do not attribute its suboptimal convergence to noise but to the method being biased.

Conversely, Table~\ref{tab:offline} shows that \ac{\OurMethod} reaches significantly higher performance than all the baselines within \numprint{10000} impressions.
Improvements are observed on all datasets under all levels of noise, especially on the commercial datasets where increases up to $0.17$ NDCG are observed.
Our results also show that \ac{\OurMethod} learns faster than the baselines; at all time-steps the offline performance of \ac{\OurMethod} is at least as good or better than all other methods,  across all datasets.
This increased learning speed can also be observed in Figure~\ref{fig:offline}.
Besides the faster learning it also appears as if \ac{\OurMethod} converges at a better optimum than \ac{DBGD} or \ac{MGD}.
However, Figure~\ref{fig:offline} reveals that \ac{DBGD} does not fully converge within \numprint{30000} iterations.
Therefore, we performed an additional experiment where \ac{\OurMethod} and \ac{DBGD} optimize models over \numprint{1000000} impressions on the MSLR-WEB10k dataset, as displayed in Figure~\ref{fig:long}.
Clearly the performance of \ac{DBGD} plateaus at a considerably lower level than that of \ac{\OurMethod}.
Therefore, we conclude that \ac{\OurMethod} indeed has an improved point of final convergence compared to \ac{DBGD} and \ac{MGD}.

Finally, Figure~\ref{fig:offline}~and~\ref{fig:long} also shows the behavior predicted by the speed-quality tradeoff~\cite{oosterhuis2017balancing}: a more complex model will have a worse initial performance but a better final convergence.
Here, we see that depending on the level of interaction noise the neural model requires \numprint{3000} to \numprint{20000} iterations to match the performance of a linear model.
However, in the long run the neural model does converge at a significantly  better point of convergence.
Thus, we conclude that \ac{\OurMethod} is capable of effectively optimizing different kinds of models in terms of offline performance.

In conclusion, our results show that \ac{\OurMethod} learns faster than existing \ac{OLTR} methods while also converging at significantly better levels of performance.

\begin{table*}[t]
\centering
\caption{Online performance (Discounted Cumulative NDCG, Section~\ref{sec:experiments:evaluation}) for different instantiations of CCM (Table~\ref{tab:clickmodels}). The standard deviation is shown in brackets, bold values indicate the highest performance per dataset and click model, significant improvements and losses over the \acs{DBGD}, \acs{MGD} and pairwise baselines are indicated by  \enkelop\ (p $<$ 0.05) and \dubbelop\ (p $<$ 0.01) and by \enkelneer\ and \dubbelneer, respectively.}
\begin{tabular*}{\textwidth}{@{\extracolsep{\fill} } l  l l l l l  }
\toprule
 & { \small \textbf{MQ2007}}  & { \small \textbf{MQ2008}}  & { \small \textbf{MSLR-WEB10k}}  & { \small \textbf{Yahoo}}  & { \small \textbf{istella}} \\
\midrule
& \multicolumn{5}{c}{\textit{perfect}} \\
\midrule
DBGD (linear) & 675.7 {\tiny (21.8)} & 843.6 {\tiny (40.8)} & 533.6 {\tiny (15.6)} & 1159.3 {\tiny (31.6)} & 589.9 {\tiny (19.2)} \\
DBGD (neural) & 602.7 {\tiny (58.1)} & 776.9 {\tiny (67.4)} & 481.2 {\tiny (53.0)} & 1135.7 {\tiny (41.3)} & 494.3 {\tiny (60.5)} \\
MGD (linear) & 689.6 {\tiny (15.3)} & 858.6 {\tiny (40.6)} & 558.7 {\tiny (6.4)} & 1203.9 {\tiny (9.9)} & 670.9 {\tiny (8.6)} \\
Pairwise (linear) & 458.4 {\tiny (13.3)} & 616.6 {\tiny (25.8)} & 345.3 {\tiny (4.6)} & 1027.2 {\tiny (9.2)} & 64.5 {\tiny (2.1)} \\
PDGD (linear) & \bf 797.3 {\tiny (17.3)} {\tiny \dubbelop} {\tiny \dubbelop} {\tiny \dubbelop} {\tiny \dubbelop} & \bf 959.7 {\tiny (43.4)} {\tiny \dubbelop} {\tiny \dubbelop} {\tiny \dubbelop} {\tiny \dubbelop} & \bf 691.4 {\tiny (12.3)} {\tiny \dubbelop} {\tiny \dubbelop} {\tiny \dubbelop} {\tiny \dubbelop} & \bf 1360.3 {\tiny (10.8)} {\tiny \dubbelop} {\tiny \dubbelop} {\tiny \dubbelop} {\tiny \dubbelop} & \bf 957.5 {\tiny (9.4)} {\tiny \dubbelop} {\tiny \dubbelop} {\tiny \dubbelop} {\tiny \dubbelop} \\
PDGD (neural) & 743.7 {\tiny (18.8)} {\tiny \dubbelop} {\tiny \dubbelop} {\tiny \dubbelop} {\tiny \dubbelop} & 925.4 {\tiny (43.3)} {\tiny \dubbelop} {\tiny \dubbelop} {\tiny \dubbelop} {\tiny \dubbelop} & 619.2 {\tiny (13.6)} {\tiny \dubbelop} {\tiny \dubbelop} {\tiny \dubbelop} {\tiny \dubbelop} & 1319.6 {\tiny (10.1)} {\tiny \dubbelop} {\tiny \dubbelop} {\tiny \dubbelop} {\tiny \dubbelop} & 834.0 {\tiny (22.2)} {\tiny \dubbelop} {\tiny \dubbelop} {\tiny \dubbelop} {\tiny \dubbelop} \\
\midrule
& \multicolumn{5}{c}{\textit{navigational}} \\
\midrule
DBGD (linear) & 638.6 {\tiny (29.7)} & 816.9 {\tiny (42.0)} & 508.2 {\tiny (21.6)} & 1129.9 {\tiny (32.2)} & 538.2 {\tiny (29.0)} \\
DBGD (neural) & 573.7 {\tiny (68.4)} & 740.3 {\tiny (69.7)} & 465.8 {\tiny (52.0)} & 1116.0 {\tiny (45.7)} & 414.3 {\tiny (96.2)} \\
MGD (linear) & 635.9 {\tiny (14.7)} & 824.5 {\tiny (34.0)} & 538.1 {\tiny (7.6)} & 1181.7 {\tiny (20.0)} & 593.2 {\tiny (9.7)} \\
Pairwise (linear) & 459.9 {\tiny (12.9)} & 618.6 {\tiny (25.2)} & 347.3 {\tiny (5.4)} & 1031.2 {\tiny (9.0)} & 72.6 {\tiny (2.2)} \\
PDGD (linear) & \bf 703.0 {\tiny (17.9)} {\tiny \dubbelop} {\tiny \dubbelop} {\tiny \dubbelop} {\tiny \dubbelop} & \bf 903.1 {\tiny (40.7)} {\tiny \dubbelop} {\tiny \dubbelop} {\tiny \dubbelop} {\tiny \dubbelop} & \bf 578.1 {\tiny (16.0)} {\tiny \dubbelop} {\tiny \dubbelop} {\tiny \dubbelop} {\tiny \dubbelop} & \bf 1298.4 {\tiny (33.4)} {\tiny \dubbelop} {\tiny \dubbelop} {\tiny \dubbelop} {\tiny \dubbelop} & \bf 704.1 {\tiny (33.5)} {\tiny \dubbelop} {\tiny \dubbelop} {\tiny \dubbelop} {\tiny \dubbelop} \\
PDGD (neural) & 560.9 {\tiny (14.6)} {\tiny \dubbelneer} {\tiny \enkelneer} {\tiny \dubbelneer} {\tiny \dubbelop} & 788.7 {\tiny (38.5)} {\tiny \dubbelneer} {\tiny \dubbelop} {\tiny \dubbelneer} {\tiny \dubbelop} & 448.1 {\tiny (12.3)} {\tiny \dubbelneer} {\tiny \dubbelneer} {\tiny \dubbelneer} {\tiny \dubbelop} & 1176.1 {\tiny (17.0)} {\tiny \dubbelop} {\tiny \dubbelop} {\tiny \enkelneer} {\tiny \dubbelop} & 390.2 {\tiny (35.1)} {\tiny \dubbelneer} {\tiny \dubbelneer} {\tiny \dubbelneer} {\tiny \dubbelop} \\
\midrule
& \multicolumn{5}{c}{\textit{informational}} \\
\midrule
DBGD (linear) & 584.2 {\tiny (41.1)} & 757.4 {\tiny (56.9)} & 477.2 {\tiny (32.2)} & 1110.0 {\tiny (37.0)} & 436.8 {\tiny (57.4)} \\
DBGD (neural) & 550.8 {\tiny (75.7)} & 720.9 {\tiny (79.0)} & 444.7 {\tiny (60.9)} & 1091.2 {\tiny (48.6)} & 322.9 {\tiny (121.0)} \\
MGD (linear) & 618.8 {\tiny (21.7)} & 815.1 {\tiny (44.5)} & 540.0 {\tiny (7.7)} & 1159.1 {\tiny (40.0)} & 581.8 {\tiny (10.7)} \\
Pairwise (linear) & 462.6 {\tiny (14.4)} & 619.6 {\tiny (25.0)} & 349.7 {\tiny (6.6)} & 1034.1 {\tiny (9.0)} & 77.0 {\tiny (2.4)} \\
PDGD (linear) & \bf 704.8 {\tiny (30.5)} {\tiny \dubbelop} {\tiny \dubbelop} {\tiny \dubbelop} {\tiny \dubbelop} & \bf 907.9 {\tiny (42.0)} {\tiny \dubbelop} {\tiny \dubbelop} {\tiny \dubbelop} {\tiny \dubbelop} & \bf 567.3 {\tiny (36.5)} {\tiny \dubbelop} {\tiny \dubbelop} {\tiny \dubbelop} {\tiny \dubbelop} & \bf 1266.7 {\tiny (50.0)} {\tiny \dubbelop} {\tiny \dubbelop} {\tiny \dubbelop} {\tiny \dubbelop} & \bf 731.5 {\tiny (80.0)} {\tiny \dubbelop} {\tiny \dubbelop} {\tiny \dubbelop} {\tiny \dubbelop} \\
PDGD (neural) & 594.6 {\tiny (23.0)} {\tiny \enkelop} {\tiny \dubbelop} {\tiny \dubbelneer} {\tiny \dubbelop} & 818.3 {\tiny (39.6)} {\tiny \dubbelop} {\tiny \dubbelop} \hphantom{\tiny \dubbelneer} {\tiny \dubbelop} & 470.1 {\tiny (19.4)} {\tiny \enkelneer} {\tiny \dubbelop} {\tiny \dubbelneer} {\tiny \dubbelop} & 1178.1 {\tiny (22.8)} {\tiny \dubbelop} {\tiny \dubbelop} {\tiny \dubbelop} {\tiny \dubbelop} & 484.3 {\tiny (64.8)} {\tiny \dubbelop} {\tiny \dubbelop} {\tiny \dubbelneer} {\tiny \dubbelop} \\
\bottomrule
\end{tabular*}

\label{tab:online}
\end{table*}

\subsection{User experience during training}
\label{sec:results:onlineperformance}

Besides the ranking models learned by the \ac{OLTR} methods, we also consider the user experience during optimization.
Table~\ref{tab:online} shows that the online performance of \ac{DBGD} and \ac{MGD} are close to each other; \ac{MGD} has a higher online performance due to its faster learning speed~\cite{oosterhuis2016probabilistic, schuth2016mgd}.
In contrast, the pairwise baseline has a substantially lower online performance in all cases.
Because Figure~\ref{fig:offline} shows that the learning speed of the pairwise baseline sometimes matches that of \ac{DBGD} and \ac{MGD}, we attribute this difference to the exploration strategy it uses.
Namely, the random insertion of uniformly sampled documents by this baseline appears to have a strong negative effect on the user experience.

The linear model optimized by \ac{\OurMethod} has significant improvements over all baseline methods on all datasets and under all click models.
This improvement indicates that the exploration of \ac{\OurMethod}, which uses a distribution over documents, does not lead to a worse user experience.
In conclusion, \ac{\OurMethod} provides a considerably better user experience than all existing methods.

Finally, we also discuss the performance of the neural models optimized by \ac{\OurMethod} and \ac{DBGD}.
This model has both significant increases and decreases in online performance varying per dataset and amount of interaction noise.
The decrease in user experience is predicted by the speed-quality tradeoff~\cite{oosterhuis2017balancing}, as Figure~\ref{fig:offline} also shows, the neural model has a slower learning speed leading to a worse initial user experience.
A solution to this tradeoff has been proposed by \citet{oosterhuis2017balancing}, which optimizes a cascade of models.
In this case, the cascade could combine the user experience of the linear model with the final convergence of the neural model, providing the best of both worlds.

\subsection{Improvements of \acs{\OurMethod}}

After having discussed the offline and online performance of \ac{\OurMethod}, we will now answer \ref{rq:performance} and \ref{rq:nonlinear}.

First, concerning \ref{rq:performance} (whether \ac{\OurMethod} performs significantly better than \ac{MGD}), the results of our experiments show that models optimized with \ac{\OurMethod} learn faster and converge at better optima than \ac{MGD}, \ac{DBGD}, and the pairwise baseline, regardless of dataset or level of interaction noise.
Moreover, the level of performance reached with \ac{\OurMethod} is significantly higher than the final convergence of any other method.
Thus, even in the long run \ac{DBGD} and \ac{MGD} are incapable of reaching the offline performance of \ac{\OurMethod}.
Additionally, the online performance of a linear model optimized with \ac{\OurMethod} is significantly better across all datasets and user models.
Therefore, we answer \ref{rq:performance} positively: \ac{\OurMethod} outperforms existing methods both in terms of model convergence and user experience during learning.

Then, with regards to \ref{rq:nonlinear} (whether \ac{\OurMethod} can effectively optimize different types of models), in our experiments we have successfully optimized models from two families: linear models and neural networks.
Both models reach a significantly higher level of performance of model convergence than previous \ac{OLTR} methods, across all datasets and degrees of interaction noise.
As expected, the simpler linear model has a better initial user experience, while the more complex neural model has a better point of convergence.
In conclusion, we answer \ref{rq:nonlinear} positively: \ac{\OurMethod} is applicable to different ranking models and effective for both linear and non-linear models.


\section{Conclusion}
\label{sec:conclusion}

In this paper, we have introduced a novel \ac{OLTR} method: \ac{\OurMethod} that estimates its gradient using inferred pairwise document preferences.
In contrast with previous \ac{OLTR} approaches \ac{\OurMethod} does not rely on online evaluation to update its model.
Instead after each user interaction it infers preferences between document pairs.
Subsequently, it constructs a pairwise gradient that updates the ranking model according to these preferences.

We have proven that this gradient is unbiased w.r.t.\ user preferences, that is, if there is a preference between a document pair, then in expectation the gradient will update the model to meet this preference.
Furthermore, our experimental results show that \ac{\OurMethod} learns faster and converges at a higher performance level than existing \ac{OLTR} methods.
Thus, it provides better performance in the short and long term, leading to an improved user experience during training as well.
On top of that, \ac{\OurMethod} is also applicable to any differentiable ranking model, in our experiments a linear and a neural network were optimized effectively.
Both reached significant improvements over \ac{DBGD} and \ac{MGD} in performance at convergence.
In conclusion, the novel unbiased \ac{\OurMethod} algorithm provides better performance than existing methods in terms of convergence and user experience.
Unlike the previous state-of-the-art, it can be applied to any differentiable ranking model.

Future research could consider the regret bounds of \ac{\OurMethod}; these could give further insights into why it outperforms \ac{DBGD} based methods.
Furthermore, while we proved the unbiasedness of our method w.r.t.\ document pair preferences, the expected gradient weighs document pairs differently.
Offline \ac{LTR} methods like Lambda\-MART~\cite{burges2010ranknet} use a weighted pairwise loss to create a listwise method that directly optimizes \ac{IR} metrics.
However, in the online setting there is no metric that is directly optimized. 
Instead, future work could see if different weighing approaches are more in line with user preferences.
Another obvious avenue for future research is to explore the effectiveness of different ranking models in the online setting.
There is a large collection of research in ranking models in offline \ac{LTR}, with the introduction of \ac{\OurMethod} such an extensive exploration in models is now also possible in \ac{OLTR}.

\subsection*{Code}
To facilitate reproducibility of the results in this paper, we are sharing the code used to run the experiments in this paper at \\ \url{https://github.com/HarrieO/OnlineLearningToRank}.
\vspace{-0.3\baselineskip}
\subsection*{Acknowledgements}
This research was partially supported by the Netherlands Organisation for Scientific Research (NWO) under project nr.\ 612.\-001.\-551.

%


\bibliographystyle{ACM-Reference-Format}
\bibliography{cikm2018-differentiable-oltr}

\end{document}